# Limits in point to point resolution of MOS based pixels detector arrays

Nicolas Fourches[a,1], D. Desforge[a], M. Kebbiri[a], V. Kumar[b], Y. Serruys[c], G. Gutierrez[c], F. Leprêtre[c], and F. Jomard[d]

[a] *CEA IRFU/DEDIP, Université Paris-Saclay*
*CEA Saclay 91191 Gif/Yvette, France*
[b] *CIMAP, 14000, Caen France*
[c] *CEA/DEN/Jannus, Université Paris-Saclay*
*CEA Saclay, 91191 Gif/Yvette, France*
[d] *GEMAC, Université de Versailles-St.Q.Y., Université Paris-Saclay, 55 avenue de Paris 78000, Versailles*
*E-mail*: nicolas.fourches@cea.fr

ABSTRACT: In high energy physics point to point resolution is a key prerequisite for particle detector pixel arrays. Current and future experiments require the development of inner-detectors able to resolve the tracks of particles down to the micron range. Present-day technologies, although not fully implemented in actual detectors can reach a 5 µm limit, based on statistical measurements, with a pixel-pitch in the 10 µm range. Attempts to design small pixels based on SOI technology will be briefly recalled here. This paper is devoted to the evaluation of the building blocks with regard to their use in pixel arrays for the accurate tracking of the charged particles. We will make here a simulations based quantitative evaluation of the physical limits in the pixel size. A design based on CMOS compatible technologies that allows a reduction of the pixel size down to the submicron range is introduced. Its physical principle relies on a buried carrier-localizing collecting gate. The fabrication process needed by this pixel design can be based on existing process steps used in silicon microelectronics. The pixel characteristics will be discussed as well as the design of pixel arrays. The existing bottlenecks and how to overcome them will be discussed in the light of recent ion implantation and material characterization.

---

[1] Corresponding author.

# Contents

## 1. Motivations for a monolithic pixel down-scalable in overall size

Particle physics motivations for the next to come collider experiments are closely linked to Higgs physics in a high energy range and the test of extra dimensions models. These future e-e+ colliders (for instance the International Linear Collider) will give a fairly good initial particle state at the Interaction Point (IP) (> 80% polarized electrons) and will exhibit very small diameter-beam size (in the order of 1 µm, 5.7 nm height and 500 nm width with 300 µm in z (z is the coordinate parallel to the beam) and make possible a very accurate determination of vertices of particle tracks at high transverse momentum. For the ILC the beam consists of trains of 2625 bunches separated by a 100 few ns [1]. This facility will provide the opportunity to make precision lifetime measurements of short lived charged particles such as b-quarks and taus. The search for extra-dimensions [2] would benefit from such a detector as it would enable precise determination of tracks, vertices and momentum of charged particles. These researches will require an inner detector which should enable to distinguish close charged tracks. Up to now a significant advance has been made with the use of monolithic pixels based on CMOS technology at the most favourable technological node. The point to point spatial resolution is key to precisely determine primary and secondary vertices and should now be expected to reach submicron resolution range. Limiting the detector degradation due to ionizing and non-ionizing radiation is still a challenge especially for hadrons colliders such as LHC. A fast electrical signal response to impinging particles is also a prerequisite. The possibility of introducing some in-pixel memory to cope with a delayed pixel-array readout should also be considered. With this implemented a triggered readout would be possible allowing the processing of only physically worthwhile data. This feature is also key to data flow reduction. The reduction in pixel size also allows a drastic reduction of the pixel hit occupancy. Hence in ILC experiments the average pixel occupancy could be down to a few hits per second. Given that the design introduced here make the pixel reset within a few tens of microsecond's this makes the hard resetting with a reset row somehow needless.

**1.1 Consequences of pixel size reduction**

A single pixel has a finite volume and can be reduced in size in its three spatial dimensions. If we consider the z direction being parallel to the beam, pixels will be set into arrays which will be arranged along the z direction for a length greater than 10 cm and a width that could reach 1 cm for instance. These rectangular shaped silicon arrays may be made from a large silicon wafer with thinning being made at the final fabrication step. The shape of the arrays will result in the inner layer being similar to a regular convex polygon [ref] in a cut-view perpendicular to the beam axis. Moreover the edges of each rectangular silicon pixel array along the z dimension will be used for the column readout and the data compression needed for fast data output to the processing electronics. For a given beam energy if we consider that the luminosity is equal to L at the interaction point (IP), the event-rate E is then proportional to L. The event rate is being here defined by the number of secondary particles generated at the IP per unit time If a hermitical detector is used the number of hits per unit time is proportional to the event rate and that is also true for charged particles. Subsequently that remains true for part of the detector such as the VXD (vertex detector) and its layers. With this in mind the number of hits per unit time depends on the area of the detector-layer and not on the number of pixels in the layer. That remains true when we normalise to a single bunch crossing. As a first conclusion we notice that the data generated during each bunch crossing does not depend on the pixel dimension, but depends on the number of hits per unit area, which in turn depends on the beam luminosity of the machine. Of course this is true below a pixel-size threshold above which there would be a number of multiple hits in a single pixel (between readout).

Table 1: size of the pixels, area, number of hit per unit area (N), address length for a pixel, data flow

| Size (lateral dimensions) | Resolution (first order binary) | Area | Number of pixels Np in an array | Number of hits per unit area and per second | Address length in bits L=log(Np)/log(2) | Data flow in bits/second |
|---|---|---|---|---|---|---|
| 1x1 µm x µm | ~ 1 µm | 10 cm squared | $10^9$ | N | Approximately 30 | 30 x N |
| 10x1 µm x µm | ~ 3 µm | 10 cm squared | $10^8$ | N | Approximately 27 | 27 x N |
| 10x10 µm x µm | ~ 10µm | 10 cm squared | $10^7$ | N | Approximately 24 | 24 x N |

Decreasing the pixel size leads to a larger number of pixels not being hit at each bunch crossing and the reduction of multiple hits in single pixels. If Sp is the pixel area and Sa the array area, the number of pixels is N= Sa/Sp., for each bunch crossing there are n hits on the array, so approximately n pixels hit. This value does not depend on N and there is a n/N proportion of pixels hit. The data flow should only depends on n. In other words data compression for each array is needed to eliminate 0 hit-pixels (not-hit pixels). In this case only the addresses of the n hit pixels (n being the number of hit pixels) then have to be output outside of the detector. This induces an increase in the data flow as the binary coded addresses are longer. The results are table above (Table 1).

This shows that an increase of resolution in binary mode binary pixels, indicating that this is a valid option [3-4] if we only output the hit pixels addresses, resulting in data flow reduction.

**1.2 Reducing single device size**

The reduction of pixel size can be achieved following two design rules (design strategies). It seems interesting to reduce the on-pixel local electronic size. This electronics exits in the so called in smart pixels which have been proposed in the past years. A local memory is often implemented. In the pixel design we introduce, this will be replaced by a memory effect that is due to a natural slow reset of the device. We have also found the way to replace the elementary 3T pixel design used in CMOS sensors by a 1T design (although some earlier studies have introduced 1T pixels usable for visible light imaging, but seemingly with limited success). This lead us to propose a new structure called TRAMOS [TRApping gate MOS]. This means that the device has a buried gate below the channel which retains the carriers that have migrated towards the buried gate. Two technologies were proposed one based on deep levels localized in the buried gate and the other with buried gate made with quantum well/dot in the valence band.

Although having some similarity with our TRAMOS pixels, DEPFETs pixels (DEPFETs) which have been developed during more than 3 decades use a doped buried gate to modulate the channel of a JFET transistor [5]. Given the dimensions of such a device, it requires a relatively thick active size (detecting volume) in order to generate a sufficient amount of electron-hole pairs, given the low sensitivity of the device. In addition as it uses a JFET which is it is not down-scalable along with its spatial dimensions. This is a drawback compared to the MOSFETs used on up-to-date silicon processes (although p-channel DEPMOS has been proposed by the DEPFET group). In spite of this, the presence of a buried gate that can effectively localize carriers during a time long enough so that a delayed readout can be used, should favour the TRAMOS structure with the advantage of being much low in the vertical dimension allowing the use of thin silicon substrates with a reduction of particle multiple scattering.

Though, in a recent study [6] GEANT4 was used to determine the distribution of the hits for high energy 130 GeV charged particles in a silicon substrate. The results [Table 2] show that for thick pixels (10 µm) most of the tracks are distributed in an area which is less than 1 µm squared.

Table 2: summary of the results: two first rows GEANT4 simulations, the last row corresponds to experimental results obtained on CMOS MAPS.

| .Thickness of the substrate | Distribution in the area | Detection efficiency |
|---|---|---|
| 10 µm | < 1µm squared | 99 % |
| 10µ-100 µm | >=2 µm | ➢ 99% |
| 5 µm | << 0.5 µm | 98 % 'from measurements |

This is compatible with realistic material thicknesses (100 µm to 10 µm). This shows that some thinning is necessary to limit the charge spread due to the hit distribution. This spread is mostly due to δ electrons and increases along with the thickness. In the 10µm-100µm range there is more charge collected in the central pixel than in the all the 8-neighbouring pixels put together (cluster of 9 pixels). To detect inclined tracks the aspect ratio (thickness/surface) should be lowered. The size (surface) can then be must be reduced along the thickness. Below10 µm in thickness the size of the pixel can be reduced to the micron squared scale. A near 99% Detection Efficiency (DE) is still obtained using simulations. Note that ideally the pixel aspect ratio should be set close to one to enable adequate inclined track detection.

To alleviate these constraints the thickness may be reduced to a few µm (5µm) at the expense of DE, the MPV (Most Probable Value) of the generated charge being of the order of 400 e-h pairs. This detection thickness is then close to that was evaluated for CMOS sensors [3-4], for which the detection efficiency (DE) measurements showed a 98 % value [3-4]. We should expect that along with CMOS sensors the noise of these new pixels will be lower than (Equivalent Noise Charge) 10-15 electrons [7]. Note that for the CMOS sensors the noise was found to be dominated by the electronic noise (temporal + FPN: Fixed Pattern Noise). With a MPS (Most Probable Signal) of 400 e the signal to noise ratio should be greater than 40, which guarantee an acceptable detection efficiency. With the TRAMOS pixel the detecting volume is fully depleted so that the electric field mitigates charge sharing on neighbouring pixels.

**1.3 Alternative designs: SOI (Silicon On Insulator) and FDSOI (Fully Depleted SOI)**

An alternative to TRAMOS would be the Fully Depleted SOI (FDSOI) with either a capacitive coupling or a DC coupling with a via down to the charge collecting electrode. Given present-days device dimensions, the buried oxide thickness can be as low as 20 nm, which is of the order of magnitude of gate oxide thickness of the 1990's DMILL process [8]. At that time (1997) a twin gated PJFET (upper control gate and lower control gate) with a buried control gate located on the top of the buried oxide was studied [9] but the device transconductance was too low with the two gates separated, this precluding its use with a via connecting the device to the collecting electrode, located in the upper side of in the bulk [9] silicon. In this design, this collecting electrode, located below the insulating buried oxide should have collected the carriers generated in the silicon substrate and used the capacitive effect on the JFET gate, with a high voltage applied to the rear bulk contact. With present day FDSOI buried oxide thicknesses, such a pixel family should be reconsidered.

**1.4 Solution proposed: technological -constraints**

Using advanced CMOS processes integrating the pixel processor at least make a pixel technology compatible with these processes is clearly an advantage.  compared with specific detector technologies used in hybrid pixel detectors for instance These hybrid pixels also use advanced CMOS processes for the readout (ROIC, readout integrated chip)  often bump-bonded to the detector. In addition these processes should be   rad-hard

The process flow necessary for the fabrication of a single TRAMOS is (planar) compatible with planar CMOS.  Two fabrication procedures are being studied. First Ge ion implantation into a silicon substrate followed by a thermal anneal , second epitaxy of a SiGe layer followed by a Si layer on a high resistivity silicon substrate with  a UHV-CVD technique. Hopefully we could make initial SiGe buried layer substrates that could accept a standard CMOS process in a following step. If this is not satisfying a CMOS process including selective ion implant or epitaxy can be the alternative. One alternative could be to use III-V materials that have better performance than IV column materials for the fabrication of quantum wells or superlattice devices. But in this case that would require the use of a fully III-V process. Such a technology could require more development time than a silicon based process.

**1.5 Solution proposed: description of the device**

Previous papers have given an accurate description of the proposed device [10-11-12-13]. We just recall here the schematic of the TRAMOS based on a SiGe quantum box acting as a buried gate (Fig.1). In a 2D cut-view, the buried layer is located 20 nm to 10 nm below the Si/Gate-Dielectric interface. Its thickness ranges from 10 nm up to 20 nm. The dimensions of the buried gate is set to 1 µm x 0.1 µm in lateral scale and 10 nm in thickness. Among other questions the buried gate thickness imposes constraints on the doping level. For this gate dimensions (referred in the previous sentence) and a $5\times10^{17}$ cm$^{-3}$ doping level this corresponds to $5\times10^7$ carriers x cm$^{-1}$ (in thickness). For a gate of 10 nm in thickness ($10^{-6}$ cm$^{1)}$) the carrier number is close to 500 not more. Hence to make the doping of the buried gate (also called Deep Trapping Gate: DTG ) effective, the doping concentration should be set slightly above $10^{16}$ cm$^{-3}$ [Table 3] which is a high value if we use un-doped silicon as initial substrates. For lower gate sizes (a factor of ten in the lateral dimensions) the doping level would be in turn close to the density of states at the top of the SiGe valence band.

Table 3: summary of results for the buried gate extrinsic properties

| Doping level (cm-3) | Thickness of the buried gate (nm) | Length of the buried gate (nm) | Width of the buried gate (nm) | Number of carriers in gate due to doping |
|---|---|---|---|---|
| $5\times10^{17}$ cm$^{-3}$ | 10 nm | 100 nm | 1000nm | 500 |
| $5\times10^{17}$ cm$^{-3}$ | 20 nm | 100 nm | 1000 nm | 1000 |

Charged particles crossing the pixel generate electron-hole pairs below the DTG Deep Trapping Gate: the buried charge collecting gate)). The DTG act as quantum box that localize holes in the valence band well and make electrons escape (due to a conduction band barrier). This makes the DTG more positively charged than in its initial state and then increase the source to drain current. In this way the structure acts as a detector.

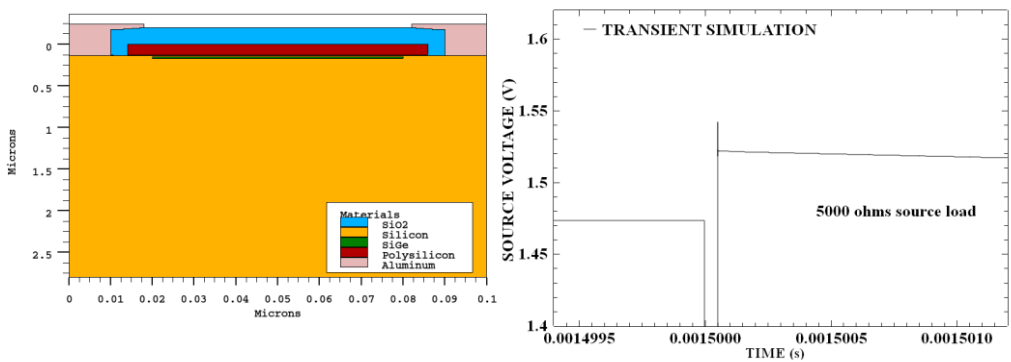

Fig. 1: 2D cut-view of the structure used for 2D device simulations (left). The thickness is reduced to 2.5 µm is this case to lower the aspect ratio (thickness/area) and for clarity.The buried gate is seen in green on the diagram close to the upper interface.. Its thickness is set to 20nm and is located 20 nm below the Si/SiO$_2$ interface. Response to an input signal of 800 e-h pairs generated in the sensitive volume (right) with a source load of 5 kohms.

The other point concerns the pixel stimuli at its input and the pixel signal at its output. The following figure (Fig.2) shows the row and column lines that is need for the proper operation of the pixel. The electrical operation of the device was described in a recent published paper [10].

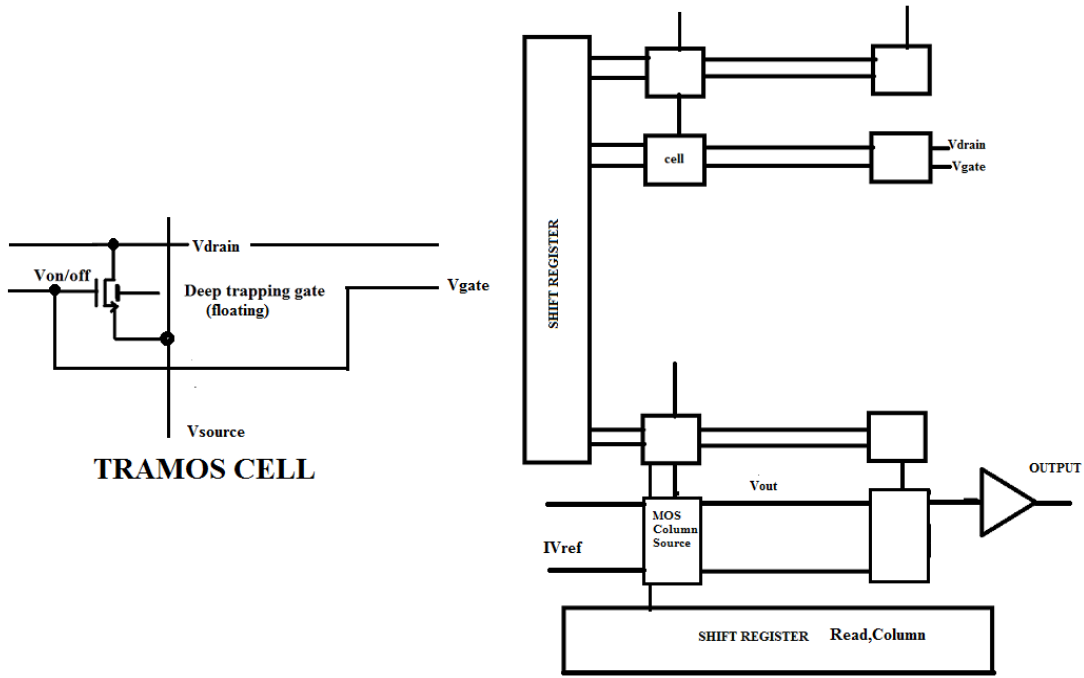

Fig. 2: The TRAMOS cell (pixel) represented her basically only requires a gate and a drain line (for each row: input pixel stimuli)) and a source line (output signal) for each column (Vsource, left). The readout scheme can be a row by row combined with column after column (right). The blank blocks are a repetition of the named block cells.

## 2. Technology and physical simulations

### 2.1 Technological guidelines : summary of results

#### 2.1.1 Zn implantation (basic and experimental study)

We have made the pixel design compatible with mainstream technologies. Among other reasons silicon remains the initial starting substrate material because it is one of the best for mechanical strength. We have made device simulations to evaluate the characteristics of the single transistor pixel. As early as 2010 [11] device simulations and technological simulations were made to assess the proof of principle for a similar pixel with a buried gate which was enriched with a high concentration of deep impurities. In this case Zn impurity was chosen despite a possible problems arising from the long-standing diffusivity of transition metals in silicon that would preclude its use in the CMOS processes. In our mind Zn could be implanted at high energy (1MeV) in the silicon substrate after oxidization or even as a post processing step so the thermal budget received by the deep Zn doped layer could be controlled. It should also avoid fast transient diffusion. It is know that Zn impurity diffuse at high temperature through interstitial (Si and Zn) mechanisms. To be quantitative at 870°C the diffusivity (diffusion constant) of Zn in dislocation rich Si is approximately equal to D ~$4\times10^{-9}$cm$^2$s$^{-1}$.

For a Zn concentration gradient of $10^{18}$ cm$^{-3}$/$10^{-7}$ =$10^{25}$ cm$^{-4}$ (concentration cm$^{-3}$/thickness cm), the number of Zn atom per second is equal to: $4\times10^{16}$ cm$^{-2}$s$^{-1}$. For one hour thermal anneal this gives : $4\times10^{16}$ x$3.6\times10^{3}$ = $1.44\times10^{20}$ Zn atoms cm$^{-2}$, this huge value shows that the thermal budget should be limited to much lower values if Zn is to be used as a deep trap.

The thermal anneal post Zn implantation should be restricted to temperature much below 870 °C. This imposes a very tough condition in the CMOS process flow. Rapid thermal anneal should be studied as an alternative to standard thermal anneal.

With no anneal and after a room temperature storage of more than six months the Zn implantation profile determined with SIMS (Secondary Ion Mass Spectroscopy) is Gaussian (Fig.3) and was found to fit with SRIM (Stopping Ranges of Ions in Matter) simulations (using both Kinchin and Pease model and full displacement cascade conditions). RBS (Rutherford Backscattering) was used to check the results and gave similar quantitative results.

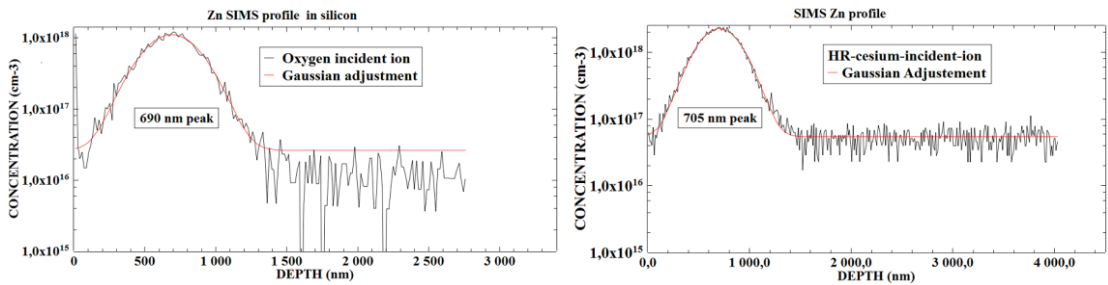

Fig. 3: concentration profile of Zn in a silicon sample obtained by SIMS. A Gaussian fit was made on all the samples: (left) high resistivity with incident oxygen ions ;( right) high resistivity with incident caesium ions

### 2.1.2 Ge deposition and implantation

The buried gate can be changed from a deep level based trapping principle to a quantum well (QW) principle. Hence as we have shown that a SiGe based QW is and effective way of implementing the DTG (as defined previously) concept [10]. What is now investigated is the process that we can used to manufacture the device. This will be based on previous knowledge on mainstream CMOS/HBT SiGe technologies. This technological background make the device fabrication feasible. Up to now appropriate technological steps are being tested. Ge ions implantation has been made on (100) silicon samples at 1MeV energy. As for the Zn ions, with no anneal and after a room temperature storage of more than six months the Ge implantation profile obtained using SIMS was found to be Gaussian and to fit with SRIM simulations results. One difficulty is that the peak concentration is lower than expected for the total dose reached. This can be alleviated by the use of higher implantation doses or successive implants/anneals. Some high temperature anneals are currently under way to get further information about the diffusivity of the Ge atoms in the silicon matrix. Results obtained with process simulations show that with a limited thermal budget, the diffusion of Ge into the upper Silicon matrix can be controlled. These results show that the use of SiGe either made by UHV/CVD or by ion implantation can be favourably considered for this pixel design.

## 2.2 Defect control and TCAD simulations

TCAD process simulations are reliable for the design of standard CMOS processes. But for these structures there are some bottlenecks due to defects control in the buried layer and on the upper n-channel. If we consider that the SiGe is the best choice for the buried gate, what is most difficult in pixel array fabrication is that implantation or deposition of Ge (or SiGe in the epitaxial case) onto Si may introduce defects into the stacked layers. Due to the lattice mismatch SiGe should be kept as a stained layer so that lattice relaxation is avoided. Lattice relaxation generates defects (mainly dislocations and defect clusters) that could render a single device dysfunctional. This sets limits on the SiGe thickness and on the layer composition (Fig. 4 of reference [10]). A trade-off of 50-75 % Ge in silicon substrate is found to be best as well as a 10 nm-20 nm thickness for the buried gate. For characterization purposes 1MeV Ge implantation in silicon samples were made too. With this energy the buried SiGe layer is located far below the surface and can be easily characterized. The results are summarized in Table 4.

Table 4: 1 MeV ion implantation in (100) silicon, the depth of the peak is ~ 600 nm. With this technique, etching and growing a Si layer will lead to Si/SiGe/Si structure which is needed for this pixel family.

| Samples | Measured (SIMS) implanted dose Integration of the distribution And Peak position | Ge peak concentration (SIMS) Sample : Cs- incident ions Negative secondary ions | RBS, Peak Ge position and concentration |
|---|---|---|---|
| C1(low resistivity) | $1,25 \times 10^{17}$ cm$^{-2}$ , 647 nm | 6% | |
| C2 (low resistivity) | $1.08 \times 10^{17}$ cm$^{-2}$, 647 nm | 5% | |
| C3 (high resistivity) | $8,49 \times 10^{16}$ cm$^{-2}$ , 658 nm | 4% | |
| C5 (high resistivity) | $8,64 \times 10^{16}$ cm$^{-2}$ , 658 nm | 4% | |
| Low resistivity sample | | | 9% ~600nm |
| High resistivity sample | | | 7% ~600 nm |

We have made Raman scattering experiments on the implanted silicon substrates. These results show that in the case of ions implantation the concentration of defects is very high in the 0-600 nm region, at the implanted dose of $10^{17}$ Ge cm$^{-2}$. Fig.4 show a comparison between as implanted 1MeV Ge and annealed sample on the not-implanted region and implanted region. This shows that the thermal treatment is sufficient to anneal out most of the defects.

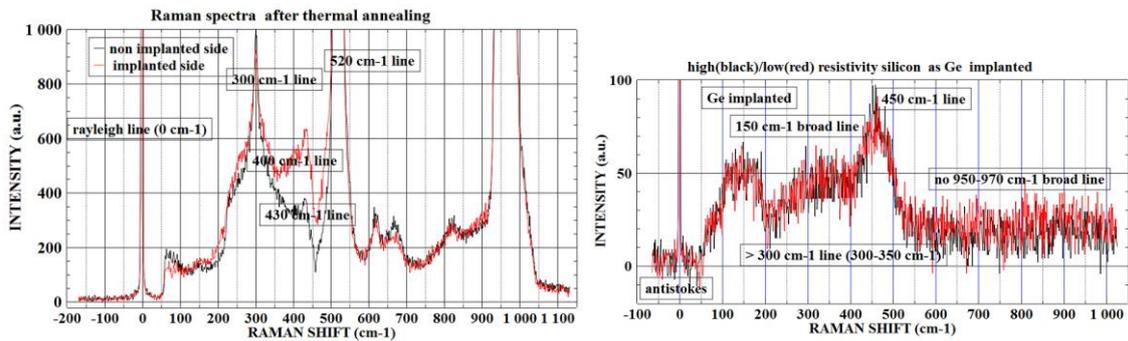

Fig. 4: Raman spectra (Stokes mode) obtained on as grown silicon and implanted and annealed silicon (left) (850 °C 50 mins) and as 1MeV Ge implanted silicon (right) . The Si-Si modes vanish after implantation and a mode seemingly due to Ge-Ge or SiGe (300-400 cm$^{-1}$) appears in the implanted zone (right).

## 3. Conclusions and short term developments

We have demonstrated here that the limits in point to point resolution of monolithic pixel based are mainly due to physics constraints that can be easily simulated, for instance in the case of high energy charged particles. These limits are not due to technological or material processing constraints. For lower energy photons (X ray down to infrared) the spatial resolution will be not as limited as they are absorbed, without induced a charged track, inside the pixel, in a small material volume containing electron-hole pairs. For photon detection the best operating mode is the Backside Illumination (BSI). We have shown that the overall constraints depend on the dimensions of the detecting device and the number of devices in the pixel. These constraints may be overcome using a single detecting and readout device. These considerations have provided the reasons to the proposed TRAMOS structure. We have shown in this case that the fabrication steps necessary to build such a device are within the reach of today's technological processes, this being confirmed by some of our experimental "building-blocks" checking. Efforts are now focusing on finalising a process flow, from which pixel array test vehicles will be fabricated. A fabrication network will be used for this purpose.

## Acknowledgments

The authors are grateful to Charles Renard (C2N, Université Paris-Sud) and Geraldine Hallais for discussions about the UHV/CVD and a first thermal anneal and Cyril Bachelet (CSNSM/Université Paris Sud) for preliminary RBS measurements. The EMIR/Jannus network provided the ion implantations and Raman measurements. Antoine Ténart (student) provide help through 3D simulation work.